\documentclass[prb,twocolumn,amssymb,epsf] {revtex4}
\usepackage{epsfig}
\def\avg#1{\langle#1\rangle}
\def\Re{\rm{Re}}
\def\Im{\rm{Im}}
\def\be{\begin{equation}}       \def\ee{\end{equation}}
\def\bea{\begin{eqnarray}}      \def\eea{\end{eqnarray}}

\def\avg#1{\langle#1\rangle}

\begin{document}
\title{Quantum dynamics, particle delocalization and instability of Mott states: \\
the effect of fermion-boson conversion on Mott states}
\author{Fei Zhou}
\affiliation{PITP and the Department of Physics and Astronomy, 
University of British Columbia, \\
6224 Agricultural Road, Vancouver,British Columbia, Canada, V6T 1Z1
}
\author{Congjun Wu}
\affiliation{
Kavli Institute for Theoretical Physics,
University of California, Santa Barbara, CA 93106, USA}
\date{\today}

\begin{abstract}
We study the quantum dynamics of superfluids of bosons hybridized with 
Cooper pairs near Feshbach resonances and the influence of fermion-boson
conversion on Mott states.
We derive a set of equations of motion which describe novel low energy
dynamics in superfluids and obtain a new distinct branch of {\em gapped} collective 
modes in superfluids which involve anti-symmetric phase oscillations in fermionic 
and bosonic channels.
We also find that Mott states in general are unstable with respect to 
fermion-boson
conversion; particles become delocalized 
and the off-diagonal long-range order of superfluids can be developed 
when a finite conversion is present.
We further point out a possible hidden order in Mott states.
It is shown that the quantum dynamics of Fermi-Bose states 
can be characterized by either an effective coupled
$U(1)\otimes U(1)$ quantum rotor Hamiltonian in a large-N limit or
a coupled XXZ $\otimes$ XXZ spin Hamiltonian in a single-orbit limit. 
\end{abstract}
\maketitle

\section{Introduction}
The phenomenon of Feshbach resonances in ultra cold atomic gases 
has 
attracted much attention. The Zeeman-field-driven two-body resonances 
between fermion pairs in open channels and bound 
molecules provide a fascinating way to tune the scattering length between 
atoms in open channels. Remarkably, this sort of simple two-body physics 
results in extremely rich quantum many-body states in atomic vapors which 
have not
been explicitly observed in conventional solid state systems.
Indeed, by varying the two-body scattering length near 
Feshbach resonances, several groups have 
successfully achieved fermionic superfluids in a strongly interacting 
regime\cite{Regal04,Jochim03,Zwierlein03,Hulet03}. 

The superfluids near Feshbach
resonances are related to the BCS-BEC crossover studied a while
ago\cite{Eagles69,Leggett80,Nozieres85,Melo93}; this was pointed out 
by a few groups \cite{Holland01,Timmermans01,Ohashi02,Stajic04,Falco04,Ho04}.
Various efforts have been made to incorporate the two-body 
resonance between Cooper pairs (open channel) and molecules (close channel)
explicitly in the many-body Hamiltonian
and many interesting results were obtained.
Relations between the multiple-channel model and
the previous single-channel model have also been studied and 
clarified.

After all these interesting efforts, a very reasonable 
understanding has been achieved. 
Three general features of Feshbach resonances deserve emphasizing here.
The first one is that 
near a Feshbach resonance the usual Cooper 
pairing amplitude
and molecule condensate wavefunction are proportional to each other.
Particularly, phases of two components (fermionic and bosonic) 
in the many-body wave 
functions are completely locked. In most cases, it has been shown that 
molecules mediate an effective interaction between fermions. 
It also has been emphasized in various occasions that molecules can be 
integrated out and at low energies one only needs to deal with an 
effective theory of 
fermions with attractive interactions.
Indeed, in the mean field approximation the Feshbach resonance introduces an effective interaction between 
fermions, the interaction constant of
which is $\gamma_{FB}^2/(2\mu-v)$. Here $\gamma_{FB}$ is the coupling 
strength (see Eq.(1)), $v$ is the detuning energy of
molecules and $\mu$ is the chemical potential of fermions.

The second feature is the behavior of many-body states at Feshbach resonances.
It turns out that the properties of states at resonances very much depend 
on the underlying
two-body parameters. If the resonance width is very larger compared with
the Fermi energy, then at resonances the energy per particle in unit of 
the Fermi energy of
free particles is universal, independent of particle densities, or
background scattering length, or other microscopic properties of 
two-body resonances.
It is also in this limit one can establish an explicit connection 
between the two-channel 
model currently employed to study the physics near Feshbach 
resonances and the one-channel model
studied long time ago.
However, if the resonance width is very narrow, then the properties 
of many-body states 
further depend on microscopic parameters of two-body physics.

The distinction between these two limits is even more severe if one
zooms in and looks into 
the molecule fraction or the chemical potential at resonances.
This is the third general 
feature we would like to turn to. At wide resonances, the chemical 
potential of fermions is 
still of the order of the free particle Fermi energy, while the molecule 
fraction is actually
inversely proportional to the width and is very small. At narrow 
resonances, the chemical 
potential is depleted to almost zero and the molecule fraction 
is substantial.

However, in these previous approaches, three important
aspects of this phenomenon have been overlooked, and sometimes, 
miscomprehended. One is the issue of quantum phase dynamics of
bosons and fermions. If we treat molecules and atoms as independent 
bosons and fermions respectively, there are no particular reasons why 
there has 
to be only one condensate phase for two-component superfluids. 
In fact, it is natural to assume that the bosonic 
or fermionic superfluid has its own quantum dynamics.  

In fact, a critical examination of the problem suggests there should be 
two phases.
Though in the mean field approximation employed 
in most of previous works on this subject, the two phases 
are usually locked, dynamically
these two phases do have their distinct features and are 
never truly identical. The extra phase degree of freedom 
indicates an extra branch of collective modes which can have rather low 
energies in 
the limit of narrow resonances\cite{Zhou05}.
These new excitations are an analogue of small fluctuations
of a relative phase between two condensates discussed in 
Ref.\onlinecite{Leggett66}. 
It remains to be 
studied 
in details and 
to be observed experimentally. 

The second issue is related to the possibility of having a
boson-fermion mixture but with decoupled low energy dynamics. Though this possibility 
hardly exists in high dimensions, in 1D there can be a phase 
transition between the usual 
phase locked superfluid and more exotic phase unlocked states. The 
critical point is determined by a Sine-Gordon type theory. This  
was discussed in a recent preprint\cite{Sheehy05} and has received further critical examination in a 
unpublished work\cite{ZCGu05}.

The third aspect is the new quantum dynamics due to the conversion between fermions and bosons. The 
conversion actually violates the particle number
conservation of fermions and bosons 
respectively and only conserves their total number.
Although this violation plays little  
role in BEC-BCS crossover superfluids because of 
the large local density fluctuations,
it can have a vital impact 
on other many-body states close to Feshbach resonances. 
One example is the instability of certain Mott states when the number 
conservation is violated. 
The other example is the development of a  certain 
hidden order in Mott states. 
The purpose of this article is to further investigate these issues and 
explore the consequences of these observations. 
A brief discussion on some of these issues was previously presented\cite{Zhou05}.

In section II, we introduce a model to study the stability/instability of
Mott states of fermion-Boson systems. We discuss the validity of the
model and the relevance 
to the physics near Feshbach resonances.
In section III, we exam the stability of certain Mott states of bosons
when Fermion-Boson conversion is present. We show that the conversion 
leads to delocalization of particles in a Mott regime and destabilizes 
the insulating phase.
A Mott state appears to develop a finite density of states
at energies well below the Mott gap.

In section IV, we further derive an effective Hamiltonian of fermion-boson
systems in a large-$N$ limit. We also obtain the equations of motion and 
investigate the novel quantum dynamics
of fermions and bosons in this limit. General structures of collective modes 
are studied in a semiclassical approximation. In section V, we
demonstrate that the delocalization leads to superfluidity by explicitly 
showing  the development of the {\em off-diagonal long ranger order}
(ODLO). These calculations also indicate that strong repulsive 
interactions between bosons or Cooper pairs do {\em not} renormalize the 
superfluid density to zero in some limit.
In section VI, we examine the hidden order in certain Mott 
states and point out various topological excitations in Mott states.
These remain to be explored experimentally.

\section{Hamiltonian for lattice Feshbach Resonances}
The model we employ to study this subject is an M-orbit
Fermi-Bose 
Hubbard Model ({\em FBHM}). 
Consider the following general form of FBHM 
\begin{eqnarray}
H&=&H_f + H_b + H_{fb}; \nonumber \\
H_f&=&-t_f \sum_{\avg{kl}, \eta,\sigma}
( f^\dagger_{k\eta \sigma} f_{l\eta\sigma} + h.c.)
+\sum_{k, \eta, \sigma} (\epsilon_\eta -\mu) f^\dagger_{k\eta\sigma} 
f_{k\eta\sigma}
\nonumber \\
&& -\lambda \sum_{k,\eta,\xi} 
f^\dagger_{k\eta\uparrow}f^\dagger_{k\eta\downarrow}
f_{k\xi\downarrow}f_{k\xi\uparrow}
+\frac{V_f}{4} \sum_k \hat{n}_{fk} (\hat{n}_{fk}-1);
\nonumber \\
H_b&=&-t_b \sum_{\avg{kl}} (b^\dagger_k b_l +h.c.)
+\sum_k (v-2\mu)b^\dagger_kb_k \nonumber \\
&+& V_b\sum_{k} \hat{n}_{bk} (\hat{n}_{bk}-1);
\nonumber \\
H_{bf}&=& - \gamma_{FB} \sum_{k,\eta}
(b^\dagger_{k} f_{k\eta\uparrow}f_{k\eta\downarrow}+h.c.)
+V_{bf} \sum_k \hat{n}_{bk} \hat{n}_{fk}.
\label{FBHM}
\end{eqnarray}
Here $k$, $\eta$ and $\sigma$ label lattice sites, on-site orbits and
spins; $\eta=1,2,...M$, $\sigma=\uparrow,\downarrow$.
$f^\dagger_{k\eta\sigma}$ $(f_{k\eta\sigma})$ is the creation
(annihilation) operator of a fermion at site $k$, with on-site 
orbital energy $\epsilon_\eta$ and spin $\sigma$.
$b^\dagger_k$($b_k$) is the creation 
(annihilation) operator of a boson at site $k$.
For simplicity, we
assume that there is only one bosonic orbital degree of freedom at each site.
The fermion and boson number operators are, respectively, 
$\hat{n}_{bk}=b^\dagger_kb_k$,$\hat{n}_{fk}
=\sum_{\eta,\sigma}f^\dagger_{k\eta\sigma}f_{k\eta\sigma}$. 
$t_f$ and $t_b$ are hopping integrals of fermions and bosons 
respectively, and hopping
occurs over neighboring sites labeled as $\avg{kl}$.
$\mu$ is the chemical potential of fermions and $v$ is the binding energy of 
bosons which are made up of bound states of fermions. 
$\lambda$ is the attractive coupling constant in the Cooper channel which we 
assume to be much larger
than the rest of couplings. Finally, $V_f$, $V_b$ and $V_{bf}$ are
the strength of repulsive interactions between 
fermions and bosons
in the density-density channel\cite{Petrov05}(One further assumes 
$V_{b}V_f > V^2_{fb}$
to ensure that homogeneous states are stable).
We only include conversion between 
a molecule and two fermions in time-reversal doublets, and 
in Eq. 1, we choose to work with doublets of $(\eta\uparrow,\eta\downarrow)$.  
FBHMs similar to Eq.(\ref{FBHM}) were previously applied
to study Bose-Fermi mixtures in optical lattices\cite{Albus03};
most recently an FBHM with fermion-boson conversion
was generalized to study the BCS-BEC crossover in lattices\cite{Carr05}. 
In the absence of the conversion term, FBHM consists of decoupled
(attractive) Fermi-Hubbard model and Bose-Hubbard model; main properties 
of latter are known \cite{Fisher89}.

In the FBHM, the conversion is between a molecule and two fermions in
the same orbit $\eta$. This approximation correctly describes the physics near Feshbach 
resonances at least in the following three limits.

a) The high density limit where
fermions mostly occupy high energy on-site orbits. 
Generally speaking, other conversion terms are allowed and the Hamiltonian should be
\bea
H'_{bf}&=&- \gamma_{FB}(\eta,\eta^\prime) \sum_{k,\eta\eta^\prime}
b^\dagger_k f_{k\eta\uparrow} f_{k\eta^\prime\downarrow}+h.c.,
\nonumber\\
\gamma_{FB}(\eta,\eta^\prime)& \sim & \int dx \phi_0^*(x) \psi_\eta(x)
\psi_{\eta^\prime}(x),
\eea
if we assume the conversion is {\em local}.
Here $\Phi_0$ is the wavefunction for bosonic molecules, and 
$\psi_{\eta,\eta^\prime}$ are wavefunctions of $\eta,\eta^\prime$
orbits at a given lattice site. If $\Phi_0$ is approximated
as a constant, thus the selection rule yields $\eta^\prime=\eta$ and 
molecules are only converted into two fermions in 
same orbits.
In a harmonic trap where $\Phi_0$ is a Gaussian wavepacket,
one then needs to take into account fermions in different orbits
as implied and demonstrated previously\cite{Esslinger05,Busch98}.

However, if orbit $\eta$ and $\eta'$ correspond to highly excited states, 
the conversion between molecules and time-reversal doublets is  
considerably larger than other terms. This yields dominating 
contributions in the large-$N$ limit.
In this case, the form of the on-site conversion 
term approaches the form in the bulk limit; 
up to a finite size effect, the conversion is between a bosonic molecule 
and two fermionic atoms
in the same orbit because of the wavefunction orthogonality.
The Hamiltonian with the conversion between a molecule and two fermions in 
same orbits thus describes the physics 
in this limit if $M$ takes a large value and the number of fermions
per site is big.

b) The low density limit near narrow resonances when non-interacting
fermions mostly occupy the lowest orbit. In this case, one can argue
that as far as the resonance width is small 
compared to the spacing between the lowest orbit and higher orbits, 
the hybridization of molecules and atoms occurs in the lowest 
energy state; so fermions remain in the 
lowest orbit. One only needs to take into
account resonances between molecular states and fermions in the lowest orbit. 
And in this case, the fermionic sector of the Hamiltonian is equivalent to
a negative-$U$ Hubbard model if one sets $M$ to be one.
(see more discussions in section VI).

c) The low density limit with
magnetic fields not too close to wide Feshbach resonances.
The validity in this limit is justified by the following observations.
Not too close to wide resonances, again the lowest eigenstate of
two interacting fermions in
a lattice site
mostly involves two fermions in the lowest orbit and a molecular bound state.
This implies that the fermion-boson conversion should again be described 
by terms such as $b^\dagger_k f_{k\eta\downarrow}f_{k\eta\uparrow}$, 
$\eta=1$.

However, right at wide resonances, the molecule state is effectively 
hybridized not only with 
two fermions in the lowest orbit
but also with two fermions in different orbits; this 
has been correctly pointed out and 
appreciated\cite{Esslinger05,Busch98}.
Even in the low density limit where free fermions occupy the lowest 
on-site orbit, the above 
FBHM Hamiltonian 
when applied to Feshbach resonances
is indeed no longer valid from this microscopic point of 
view. It remains to be understood how many-body physics will be affected 
by this complication.

Without losing generality, in this article we study the effective low energy theory in a 
limit where the 
fermion-boson conversion strength is weak (i.e., narrow resonance) and discuss the issue 
of Mott states'
instability.
However, we would like to argue that physics discussed in this article 
would {\em not} be affected by the presence of
additional conversion terms when the conversion strength is strong (i.e., wide resonance).
The main reason is that the form of the long wave length effective 
Hamiltonian described below is subject to severe constraints from the 
symmetries and 
hydrodynamics in our problem and has little dependence on microscopics. As 
far as these extra terms only renormalize coefficients in 
equations of motion but
do not alter the general form of hydrodynamics discussed in the article,
most of conclusions arrived here remain valid even in this delicate limit.
This is evident from a general renormalization point of view but has been 
unfortunately overlooked in the last two references of 
Ref.\onlinecite{Busch98}.
We believe that the significance of extra conversion terms on
the long wave length physics has been overstated previously.

\section{Delocalization of particles under the influence of
Feshbach resonances}
In this section, we are going to study a Mott state under the influence of 
Feshbach resonances, especially the
effect of fermion-boson conversion.
A Mott state of bosons or Cooper pairs appears whenever bosons or 
Cooper pairs in lattices are strongly repulsively 
interacting and if the corresponding filling factors are integers. 
One of important properties of a Mott state is its
incompressibility, or a finite energy gap in its excitation spectra,
thus 
a Mott state is believed to be robust. 
When hopping is renormalized to zero due to repulsive interactions, the 
number of particles at each site can be strictly quantized and 
discrete; particles are {\em locally} conserved. 

Below we are going to show that in general Mott states are unstable 
with respect to Feshbach resonances. The primary reason is that 
particle numbers of fermions or bosons involved in resonating conversion are not 
conserved separately. So the conversion  not only mediates an 
attractive interaction between fermions as realized before, 
but also, more importantly violates the local conservation law. This 
introduces new low energy degrees of freedom and results in {\em a novel 
mechanism to transport particles}.  
It leads to delocalization of particles in the limit of large 
repulsive interactions.

To address the issue of localization of particles, we first
introduce the following time-ordered Green's functions\cite{Mahan00}
\begin{eqnarray}
&& G^b(t,0; k,0)= -i\langle {\cal T} b^\dagger_k(t)b_0(0) \rangle \nonumber \\
&& G^f(t,0;k,0)= -i \langle {\cal T} f^\dagger_{k\eta\sigma}(t)f^\dagger_{k\eta \bar\sigma}(t)
f_{0\eta'\sigma'}(0)f_{0\eta' \bar\sigma'}(0) \rangle.\nonumber \\
\end{eqnarray}

\begin{figure}
\centering\epsfig{file=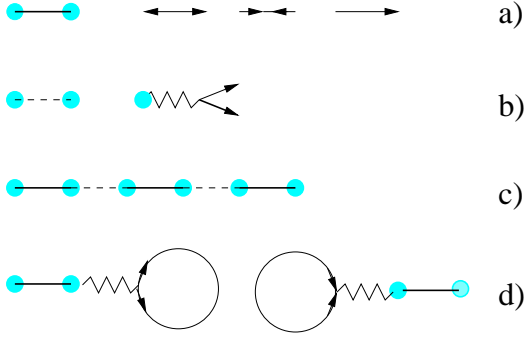,clip=1,width=0.8\linewidth,angle=0}   
\caption{Diagrams leading to the delocalization of bosons in Mott states. 
Solid lines with circles are for boson propagators while lines with 
arrows are for fermion propagators.
a) (from left to right) bosonic propagators with no hopping; 
time ordered fermionic anomalous propagators 
$ -i\langle {\cal T}f^\dagger f^\dagger\rangle$
and $-i\langle {\cal T} f f \rangle$; the fermionic normal propagators
$-i\avg{{\cal T}f^\dagger f}$.
b) (from left to right) vertices for hopping of bosons, and
for fermion-boson conversion;  
c) The contribution to the bosonic propagator at a 
large distance due to hopping.
d) The contribution to the propagator at a large distance
due to fermion-boson conversion.}
\label{fig:bsfmdiag}
\end{figure}

Now we assume $t_b/V_{b} \ll 1$ and the number of bosons per
lattice site $n_b$
is an integer so that the ground
state of bosons is a Mott state. 
Without losing generality, we also assume that the chemical potential $\mu$
is precisely in the middle of the Mott gap so that the system is 
particle-hole symmetric and $V_{bf}=0$.
Meanwhile, the number of fermion 
pairs per lattice site is either a non-integer or an integer but 
$t_f/V_f \gg 1$ so that the ground state of fermions is a superfluid. 
We are interested in the 
effect of fermion-boson conversion on the Mott state.

When there is 
 no fermion-boson conversion, one can evaluate the boson 
Green's function by an expansion in terms of the parameter $t_b/V_b$.
For instance, the zeroth order Green's function is
\begin{eqnarray}
 G^{b0}(\epsilon; k,0)&=&(\frac{n_b}{\epsilon - V_b+i\delta} 
-\frac{n_b+1}{\epsilon+V_b-i\delta})
\delta_{k,0},
\end{eqnarray}
reflecting the zero bandwidth in this limit. All low energy excitations 
are localized 
and gapped with a single energy $V_b$.
Here $n_b$ is the number of bosons per lattice site.
The small finite hopping amplitude leads to corrections to this form of 
the Green's 
function; following the diagram in Fig. 1c, 
one finds that
\begin{equation}
\delta G^{b} (\epsilon; k,0) \sim (\frac{2 t_b n_b V_b}{\epsilon^2- 
V^2_b+i\delta})^{R_k},
\label{correction}
\end{equation}
where $R_k$ is the distance between two lattice sites $k$ and $0$.
To obtain this result, we have assumed that $n_b$ is much larger than one.
It is obvious that at low energy $\epsilon \ll V_b$, the two-point 
Green's function decays exponentially as a function of distance $R_k$.
Eq.(\ref{correction}) also implies that the localization length at small finite $t_b$ 
should scale as
\begin{equation}
\xi_L \sim \ln^{-1} \frac{V_b}{n_b t_b}
\end{equation}
in the unit of the lattice constant.

The localization of particles in a conventional Mott state is largely due 
to the absence of available low energy states below the energy
scale set by 
$V_b$. So to remove a particle at the point 0 and for the particle to travel 
to the point $k$, one has to confront a sequence of energy barriers of 
height $V_b$. This blockade results in the localization.

\begin{figure}
\centering\epsfig{file=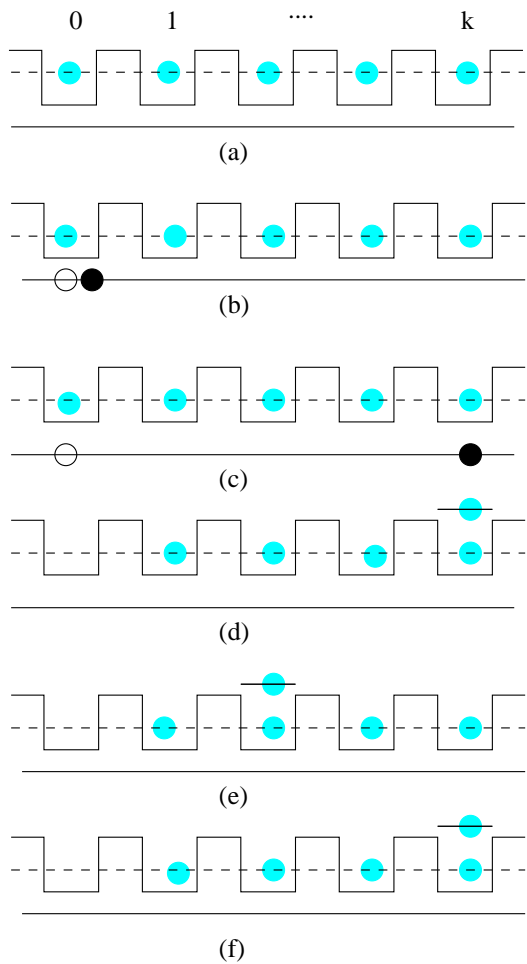,clip=1,width=0.8\linewidth,angle=0} 
\caption{ Schematic of creation of particle-hole excitations with 
(a)-(d)  and without  (e)-(f) (color online)  
fermion-boson conversion. Thick circles with light blue color
are for bosons. Thin circles for holes left by fermion pairs
and filled circles with black color for fermion pairs. 
Lines below the periodical structures are schematics of Fermi seas.
a) The ground state of bosons (in a Mott state) and fermions 
(in a superfluid state).
b) Creation of a Cooper pair and a hole pair in fermionic
 superfluid channel at site 0;
c) propagation of the Cooper pair to site $k$; d) after the 
conversion of a boson into a particle Cooper pair at site 0 and 
a particle pair into a boson at site k takes place, a final state
with one extra boson at site k and a bosonic hole at site $0$. 
The Fermi superfluid is in its ground state. In e)-f), a boson 
at site k and a bosonic hole at site 0 are created without 
the fermion-boson conversion.
Note in e), a particle effectively experiences an energy barrier 
with height $V_b$; the amplitude of finding a 
particle-hole pair separated with a large distance is 
therefore exponentially small.}
\label{fig:bsfmcon}
\end{figure}

When a fermionic superfluid is present and the fermion-boson
conversion occurs,
there is an additional channel for a particle to travel from site 0 to $k$. 
The mechanism is schematically shown in Fig. \ref{fig:bsfmcon}. 
Instead of removing a
bosonic particle from site 0 and adding to site $k$, one can remove a Cooper 
pair at site 0 and transport it to site $k$. For this, a Cooper pair 
experiences no energy barrier imposed by repulsive interactions because 
there are sufficient low energy degrees of freedom available for 
particle-hole excitations in a superfluid. At a latter stage, 
one then turns on Feshbach resonances to remove the boson at site 0
by converting it into a Cooper pair to fill up the hole left behind by 
the transported Cooper pair; similarly, the Cooper pair transported can 
be converted to a boson as an additional particle at site $k$. 
The net effect is that a bosonic hole is created at site 0 and
bosonic particle excitation is now at site $k$.
Since the fermionic channel has a long range order, this process
therefore yields a long range particle-hole excitation.

At a formal level, one can study this contribution by introducing a
vertex for the fermion-boson conversion. Furthermore, in the weakly 
coupling limit, the long range component of the fermion Green's
function reflects the usual off-diagonal-long range order.
In the mean field approximation, one obtains 
\bea
&& G^f(\epsilon; k, 0)\sim  \delta({\epsilon}) F(R_k)
+ \mbox{...}
\eea
Here $...$ represents other contributions which decay
over large distance.
Let us emphasize that $F(R_k)$ is a constant
and is independent of the distance $R_k$ between $k$ and $0$.

Following the diagrams in Fig. \ref{fig:bsfmdiag}d, one obtains the 
contribution of the boson Green's function 
\begin{equation}
\delta G^b(\epsilon;k,0)
\approx (\frac{\gamma_{FB}}{V_b})^2 G^f(\epsilon; k,0). 
\label{green3} 
\end{equation} 
Eq. (\ref{green3}) illustrates two important properties of the state 
under consideration, which are intimately connected.
Firstly, the non-exponentially decay component of the boson Green's function 
is {\em proportional} to the 
off-diagonal-long-range order in the fermionic
channel.  
The long range component in Eq.(\ref{green3}) shows that 
at least a fraction of all bosons actually become delocalized.
The fermion-boson
conversion effectively leads to the delocalization of bosons
as argued above (also see Fig.\ref{fig:bsfmcon} for more explicit
discussions); it is the delocalization of bosons which in fact induces superfluidity in 
the bosonic channel.

Secondly, the zero energy peak (at $\epsilon=0$) in the Green's function in
the mean field approximation suggests that some bosons should now 
condense at the zero energy;
the condensation fraction is $(\gamma_{FB}/V_b)^2$.
Notice that now the resultant bosonic state is compressible.
It is thus implied that there should be additional low energy states well below
the original Mott gap. These extra states are one of the consequences of the unusual hydrodynamics in the problem.
We also anticipate that the low energy structure of $\delta G^b$ such as 
the peak height is to be modified when various
fluctuations are taken into account. 

In the next two sections, we are going to analyze the long range order in 
details.
To understand the induced superfluidity, it is most convenient to 
first obtain an effective theory where the typical issues of broken 
symmetries can be easily addressed. So in section III, we derive
an effective coupled $U(1)\otimes U(1)$ quantum rotor model for the FBHM.
In section IV, we employ the effective model to examine the long
range order.

\section{An effective Hamiltonian and the equation of motion}
\subsection{a $U(1) \otimes U(1)$ coupled quantum rotor model}
We first study the {\em large-N} limit where $n_b$ and 
$n_f (< M)$, the average numbers of fermions and 
bosons are both much bigger than unity. Because $\lambda$ is much larger
than other
coupling constants, the ground state of fermions for the on-site part of the 
Hamiltonian $H_f$ naturally should be a BCS state. 
For simplicity, we also assume that the Fermi energy as well as the BCS gap are larger 
than the 
Fermi-Boson
coupling strength $\gamma_{FB}$ so that we can neglect the Fermi degrees of freedom
at low energies and obtain the effective Hamiltonian written in terms of various collective coordinates
(see below).

This suggests that it should be 
convenient to work with the following coherent state representation,
\begin{eqnarray}
&& |\{ \phi_{fk} \}; \{ \phi_{bk} \}>=
\prod_k 
\sum_{n_{bk}}
g_0(n_{bk})\frac{[\exp(-i\phi_{bk}) b_k^\dagger]^{n_{bk}}}{\sqrt{n_{bk}!}}
\nonumber \\
&& \otimes 
\prod_\eta (u_\eta + v_\eta 
\exp(-i\phi_{fk}) f^\dagger_{k\eta\uparrow} f^\dagger_{k\eta\downarrow}) 
|0 \rangle.
\label{coherent}
\end{eqnarray}
Here $u_\eta, v_\eta$ are the coherence factors in the BCS wavefunction which
minimize the total on-site energy;
$g_0(n_{bk})$ is a unity for $n_{max}+n_b > n_{bk} > n_b - n_{max}$,
$n_{max}$ is much larger than one.
These states form a low energy Hilbert subspace
and are orthogonal in the limit which interests us,
or $\avg{|\{\phi'_{fk} \}; \{ \phi'_{bk} \}|$$
\{\phi_{fk} \}; \{ \phi_{bk}\}|}$ is equal to zero if 
$\phi_{fk} \neq \phi'_{fk}$ or
$\phi_{bk} \neq \phi'_{bk}$.

At last, in the coherent-state representation one shows that
$\hat{n}_{fk}/2 =i \partial/\partial {\phi_{fk}}$,
and $\hat{n}_{bk}=i\partial/\partial {\phi_{bk}}$; or
\begin{eqnarray}
&& [\frac{1}{2}\hat{n}_{fk}, \exp(-i\phi_{fk'})]= \delta_{k,k'} 
\exp(-i\phi_{fk}),
\nonumber \\
&& [\hat{n}_{bk}, \exp(-i\phi_{bk'})]= \delta_{k,k'} \exp(-i\phi_{bk}).
\label{conjugate}
\end{eqnarray}

So in the subspace of coherent states, we find the effective Hamiltonian is
\begin{eqnarray}
&& H_{eff}=
-J_f \sum_{<kl>} \cos(\phi_{fk}-\phi_{fl})
+\frac{V'_f}{4} \sum_{k} (\hat{n}_{fk} -n_f)^2 \nonumber \\
&&
-J_b \sum_{<kl>} \cos(\phi_{bk}-\phi_{bl})
+V_b \sum_{k} (\hat{n}_{bk} -n_b)^2 \nonumber \\
&&-
\sum_{k} 
\Gamma_{FB} 
\cos(\phi_{fk}-\phi_{bk})
+V_{bf} (\hat{n}_{bk}-n_b) (\hat{n}_{fk}-n_f).  \nonumber \\
\label{effH}
\end{eqnarray}
The exchange couplings $J_f$, $J_b$ and $\Gamma_{FB}$ can be estimated
as
\begin{eqnarray}
&& J_f=t_f^2\sum_{\eta,\eta'}
\frac{u_\eta v_\eta u_\eta' v_\eta'}{E_\eta + E_\eta'}, \nonumber \\
&& J_b = n_b t_b, \Gamma_{FB}=\gamma_{FB} 
\sqrt{n_b}\sum_\eta u_\eta v_\eta;
\end{eqnarray}
$E_\eta=\sqrt{(\epsilon_\eta-\mu)^2 +\Delta_0^2}$ is the quasi-particle energy 
and $\Delta_0$ is the BCS energy gap. 
Furthermore, 

\begin{equation}
V'_f=V_f +\frac{\partial^2 {\cal E}_k (n_{fk})}{\partial n_{fk}^2}.
\end{equation}
${\cal E}_k$ is the on-site energy of $n_{fk}$ particles and 
its second derivative is inversely proportional to  
the compressibility of a BCS state. In a recent work of one of 
the authors\cite{Zhou05}, it was assumed that $V_f$ is much bigger than 
the second term in the above equation and $V'_f\approx V_f$. However,
when Fermions do not have repulsive interactions in the density-density 
channel ($V_f=0$),
$V'_f$ is equal to ${\partial \mu_{BCS}}/{\partial n_{fk}}$, $\mu_{BCS}$ is the 
chemical potential of $n_f$ fermions in a BCS state,
which is typically of order of the one-particle level spacing 
at each lattice site. 
From now on,
we will assume $V_f$ is much larger than the level spacing and omit prime in $V'_f$

$n_f$ and $n_b$ are functions of $\mu$, $v$ and $V_{f,b,fb}$:
\begin{eqnarray}
&& n_f=\frac{2V_b(\mu_0+V_f/4)-V_{bf}(2\mu-v+V_b)}{V_fV_b-V^2_{bf}}
\nonumber \\
&& n_b=\frac{V_f(2\mu-v+V_b)-V_{bf}(2\mu_0+V_f/2)}{2(V_fV_b-V^2_{bf})}.
\end{eqnarray}
Here $\mu_0=\mu-\mu_{BCS}$. 
Obviously, the detuning energy $v$ has to be sufficiently small in order for
$n_b$ to be positive. 

Eq. (\ref{conjugate}) and Eq. (\ref{effH})  define the low energy quantum
dynamics of fermions and bosons under the influence of fermion-boson
conversion in Feshbach resonances. 
In the absence of Feshbach resonances ($\Gamma_{FB}=0$) and $V_{bf}$, 
the effective Hamiltonian describes two
decoupled sets of quantum $U(1)$ rotors in a lattice, the behaviors of 
which are well known. If $n_f/2$ or $n_b$ is a positive integer, the effective 
model can be used 
to study superfluid-Mott state transitions. 
A Mott phase corresponds to $U(1)$ symmetry restored states and
$U(1)$-symmetry breaking solutions represent a superfluid phase.
For the bosonic (Cooper pair) sector, the phase transition takes place when
$r_f=zJ_f/V_f$ ($r_b= zJ_b/V_b)$ is equal to a critical value 
$r_{fc}$($r_{bc}$) ($z(>1)$ is the coordination number).
The critical values which are of order of unity are usually 
calculated numerically.

In the presence of $\Gamma_{FB}$, the Hamiltonian Eq.(\ref{effH})
describes a coupled
$U(1)\otimes U(1)$ quantum rotor model in a lattice. 
$U(1)\otimes U(1)$ symmetry breaking solutions
when both $r_{f,b}$ are much larger than unity correspond to a superfluid 
phase.

In general,
the wavefunctions for the 
many-body ground state and excitations $\Psi_n(\{ \phi_{bk} 
\}; \{ \phi_{fk} \})$($n=0,1,2,...$) are the eigenstates of the 
Hamiltonian
in Eq.(\ref{effH}). The boundary conditions are periodical along the directions
of $\phi_{fk,bk}$ with a period $2\pi$, so the wavefunctions are effectively 
defined on an 
$S^1\otimes S^1$ torus with radius of each $S^1$ equal to one. 
If the average number $n_f/2$ and $n_b$ are integers, 
one introduces a {\em gauge} transformation

\begin{eqnarray}
\Psi \rightarrow \Psi \prod_k \exp(-i n_f\phi_{fk}/2-i n_b 
\phi_{bk});
\end{eqnarray}
the shifted number operators become

\begin{eqnarray}
&& \delta\hat{n}_{fk}/2=\frac{1}{2}(\hat{n}_{fk}-n_f)=i\partial/\partial \phi_{fk},
\nonumber \\
&& \delta\hat{n}_{bk}=\hat{n}_{bk}-n_b=i\partial/\partial \phi_{bk}.
\end{eqnarray}
The effective Hamiltonian and eigenstates in the shifted basis are given by the following equation

\begin{eqnarray}
&& [- \sum_k V_f \frac{\partial^2}{\partial \phi^2_{fk}}
+ V_b \frac{\partial^2}{\partial \phi^2_{bk}}
+2 V_{bf}
\frac{\partial}{\partial \phi_{fk}}\frac{\partial}{\partial \phi_{bk}}
\nonumber \\
&& -J_f \sum_{<kl>}\cos(\phi_{fk}-\phi_{fl})
-J_b \sum_{<kl>}\cos(\phi_{bk}-\phi_{bl})
\nonumber \\
&& -\Gamma_{FB} \sum_{k}\cos(\phi_{fk}-\phi_{bk})]\Psi_n=E_n\Psi_n. 
\end{eqnarray}
It is evident, following the above equations that when both fermions and bosons are in 
superfluid
phases, quantum phases of two superfluids are locked to minimize the potential energy 
$-\Gamma_{FB} \cos(\phi_{bk}-\phi_{fk})$. That is 

\bea
\phi_{fk}=\phi_{bk}=\phi_0.
\eea

So in this new basis a spontaneous symmetry breaking solution
with the wavefunction 
\begin{eqnarray}\Psi \sim \prod_k 
\delta(\phi_{fk}-\phi_0)\delta(\phi_{bk}-\phi_0)
\end{eqnarray}
represents a typical superfluid.
A symmetry-unbroken solution with the wavefunction
\begin{eqnarray}
\Psi \sim \prod_k {(2\pi)^{-1}}\exp(im_{fk}\phi_{fk})
\otimes \exp(im_{bk}\phi_{mk})
\end{eqnarray}
($m_{fk,bk}=0$ for all $k$) on the other hand 
corresponds to
a Mott state with $ \delta\hat{n}_{fk(bk)} \Psi=0$ or 
$\hat{n}_{fk(bk)}\Psi=n_{f(b)}\Psi$
at each lattice site.

\subsection{The equations of motion and general features of
collective modes}
In a superfluid phase,
the Hamiltonian in Eq.(\ref{effH}) further leads to the following 
semiclassical equation of motion in the long wave length limit 
\begin{eqnarray}
&& \frac{\partial \phi_{fk}}{\partial t}=V_f \delta \hat{n}_{fk}
+ 2 V_{bf} \delta \hat{n}_{bk},\nonumber \\
&& \frac{\partial \phi_{bk}}{\partial t}= 2V_b \delta \hat{n}_{bk}
+ V_{bf} \delta \hat{n}_{fk},\nonumber \\
&& \frac{1}{2}\frac{\partial \delta{\hat{n}}_{fk}}{\partial t}= {J_f}
\Delta \phi_{fk}+\Gamma_{FB} (\phi_{bk}-\phi_{fk}),
\nonumber \\
&& \frac{\partial \delta{\hat{n}}_{bk}}{\partial t}={J_b}
\Delta \phi_{bk}+\Gamma_{FB} (\phi_{fk}-\phi_{bk}).
\label{eom}
\end{eqnarray}
Here $\delta\hat{n}_{fk,bk}=\hat{n}_{fk,bk}-n_{f,b}$.
we have taken a continuum limit and $k$ labels the coordinate of
phases of bosons and fermion pairs ($\phi_{fk,bk}$) in this equation; 
$\Delta$ is a Laplacian operator.
The lattice constant has been set to be one.
The above set of equations were previously derived\cite{Zhou05}.

In the absence of Fermion-boson conversion, the third and fourth 
formulae in Eq.(\ref{eom}) are the conservation laws for fermions 
and bosons respectively.
\bea
&& \frac{1}{2}\frac{\partial \delta{\hat{n}}_{fk}}{\partial t}+ 
\nabla \cdot {\bf J}_{fk}=0,
\nonumber \\
&& \frac{\partial \delta{\hat{n}}_{bk}}{\partial t}+
\nabla \cdot {\bf J}_{bk}=0,
\eea
where supercurrents are defined as ${\bf J}_{fk,bk}=-J_{f,b}
\nabla \phi_{fk,bk}$
(the definition of phases differs from the conventional one by a minus sign).
Obviously, the fermion-boson conversion violates the conservation law and introduces
a source term which is proportional to $\Gamma_{FB}$. It is this new quantum 
dynamics which yields the delocalization 
in the previous section. Below we show that in addition to the usual gapless Goldstone mode, 
the quantum dynamics in this case also leads to a new branch of 
collective modes which are fully gapped.

Let us introduce the plane wave representation for $\phi_{fk,bt}(t)$
and study the eigenmodes. 
The above semiclassical equation suggests 
spectra of collective excitations.
The equation for eigen modes $\phi_{f,b}(\omega,{\bf Q})$
reads as
\bea
&& [ \omega^2 \left( \begin{array}{cc} 
M_{ff} & M_{fb}\\
M_{bf} & M_{bb}
\end{array}
\right)
-Q^2 \left(\begin{array}{cc}
J_f & 0 \\
0 & J_b
\end{array}
\right) 
\nonumber \\
&& -\Gamma_{FB} 
\left(
\begin{array}{cc}
1 & -1 \\
-1 & 1
\end{array}
\right ) 
] 
\left(
\begin{array}{c}
\phi_f(\omega,{\bf Q}) \\
\phi_b(\omega,{\bf Q})
\end{array}
\right)=0.
\eea
Here the matrix elements $M_{\alpha,\beta}$($\alpha$,$\beta=b,f)$
are defined as
\bea
&& M_{ff}=\frac{1}{2}\frac{V_b}{V_f V_b-V^2_{bf}}, \nonumber \\
&& M_{bb}=\frac{1}{2}\frac{V_f}{V_fV_b-V^2_{fb}}, \nonumber \\
&& M_{bf}=M_{fb}=-\frac{1}{2}\frac{V_{bf}}{V_f V_b-V^2_{fb}}.
\eea
The eigenfrequencies of modes are the solutions of the
following equation

\bea
&& \frac{\omega^4}{4} -\frac{\omega^2}{2} [Q^2(J_f V_f + J_b 
V_b)+\Gamma_{FB}(V_f+V_b-2 V_{bf})]
\nonumber \\
&& +[J_f J_b Q^4 +\Gamma_{FB}(J_b +J_f) Q^2] (V_bV_f-V^2_{fb}).
\label{eigen}
\eea

By solving the equation for eigen frequencies, one obtains
the collective mode spectrum.  
The above equation shows that there should be two branches of 
collective modes the dispersion relations of which are given below:
\begin{eqnarray}
&& a) \omega^2=\alpha |{\bf Q}|^2, 
\phi_f(\omega, {\bf Q} \rightarrow 0)=\phi_b(\omega, {\bf Q} \rightarrow 
0);
\nonumber \\ 
&& b) \omega^2=\Omega_0^2+\beta |{\bf Q}|^2, \nonumber \\ 
&& \phi_f(\omega, {\bf Q} \rightarrow 0)=-
\frac{V_f-V_{bf}}{V_b-V_{bf}}
\phi_b(\omega, {\bf Q}\rightarrow 
0).
\label{dispersion}
\end{eqnarray}
$\phi_{f,b}(\omega,{\bf Q})$ are the Fourier components of phase
fields $\phi_{fk,bk}(t)$. 
It is worth emphasizing that
in the long wave length limit,
mode a) is fully symmetric in phase oscillations of fermions and bosons, 
independent of various parameters;
mode b) represents out-of-phase oscillations in fermionic and bosonic 
channels and becomes fully antisymmetry when $V_b=V_f$.
In the absence of conversion ($\Gamma_{FB}=0$),
these two modes correspond to two gapless Goldstone modes 
associated with breaking two decoupled $U(1)$ symmetries.
However, in the presence of Feshbach resonances 
only the symmetric mode a) remains gapless 
corresponding to the usual Goldstone mode of superfluid
while the antisymmetric mode b)
is fully gapped because of the phase-locking effect of Feshbach 
resonances. 

In general, $\Omega_0$, $\alpha$ and $\beta$ depend on various parameters 
in the Hamiltonian; $\Omega_0$ is always proportional to $\Gamma_{FB}$,
and $\alpha$ on the other hand is independent of $\Gamma_{FB}$.
When $V_{bf}=0$, $V_f=V_b=V_0$ and $J_f=J_b=J_0$,
Eq.\ref{eigen} becomes\\
\bea
\frac{\omega^4}{4} &-&\frac{\omega^2}{2} [2J_0V_0 Q^2+2\Gamma_{FB}V_0] + 
J_0^2 Q^4 \nonumber \\
&+&
2\Gamma_{FB}J_0V_0^2 Q^2 =0.
\eea
Consequently,
the dispersion relations are given by Eq.(\ref{dispersion}) with 
$\Omega^2_0= 4 \Gamma_{FB} V_0$,
$\alpha=\beta=2J_0V_0$. 

A more interesting and realistic limit is when $V_{bf}$ is small and set 
to be zero but $V_{b,f}$ are not equal. In this case, 
Eq.(\ref{eigen}) becomes 
\bea
&& \frac{\omega^4}{4} -\frac{\omega^2}{2} [Q^2(J_f V_f + J_b 
V_b)+\Gamma_{FB}(V_f+V_b)]
\nonumber \\
&& +[J_f J_b Q^4 +\Gamma_{FB}(J_b+J_f) Q^2] V_bV_f = 0
\label{eigen1}
\eea
One then obtains the 
dispersion with coefficients given below
\bea
&& \alpha=2(J_f+J_b) \frac{V_b V_f}{V_f +V_b}, \nonumber\\
&& \beta=2\frac{J_f V_f^2 + J_b V_b^2}{V_f + V_b}, \nonumber \\
&& \Omega_0^2=2\Gamma_{FB} (V_b+V_f).
\eea
Note that in the noninteracting limit, $V_f$ is equal to $\partial 
\mu/\partial n_f$ and is finite.
The above equations show that the velocity of mode a)
(the symmetric Goldstone mode), $\sqrt{\alpha}$, decreases when interactions between
bosons $V_b$ become smaller. This is because as bosons become weakly 
interacting, the density fluctuations in the symmetric mode are dominated 
by those of bosons and the fermion density fluctuations become 
insignificant. So although the sound velocity of fermion superfluids 
is finite, the fermionic contribution to the symmetric mode is negligible 
and the Goldstone mode becomes softer and softer as $V_b$ goes to zero.

On the other hand, the gap in the antisymmetric mode b) remains finite in 
the 
limit when $V_b=0$; that is $\Omega_0^2=2\Gamma_{FB} V_f$. 
As discussed in the previous subsection, if repulsive interactions 
between fermions are zero, $V_f$ approaches the value of 
$\partial\mu/\partial n_{bk}$.
One also notice that when $V_{bf}$ is zero, the total density fluctuations in mode b) 
at $Q=0$ are zero, that is
\bea
\delta n_{bk}=-\delta n_{fk}
\eea
following the last equation in Eq.(\ref{dispersion}).

The above semiclassical approach to collective modes is valid 
when $V_{f,b,fb}$ are small so that various renormalization 
effects can be neglected.
Collective modes in a large-V limit can be more conveniently studied using 
a saddle point expansion. This alternative approach to study the collective 
modes is explored in a unpublished work\cite{CJ05}.

\section{Development of ODLO in a large-V limit}

\subsection{Molecule mean field approximation}
We first introduce order-parameters to classify various states.
The order parameters which can be used to classify states are

\bea
\tilde{\Delta}_b=<b^\dagger_k>, 
\tilde{\Delta}_f=<f^\dagger_{k\eta\sigma}f^\dagger_{k\eta-\sigma}>.
\eea
When $\tilde{\Delta}_b$ is nonzero (zero), the ground state is a bosonic superfluid (Bosonic 
Mott state),
or SFb (MIb). When $\tilde{\Delta}_f$ is nonzero (zero), the corresponding state is a fermionic 
superfluid
(fermionic Mott state), or SFf (MIf).

Below we demonstrate the invasion of superfluidity into a parameter region 
where only Mott states are expected to be ground states if there were no 
Feshbach resonances.
To understand the influence of Feshbach 
resonances on Mott states, we first consider a situation where
again both $n_b$ and $n_f/2$ are {\em integers} and $r_b$ is much less than 
$r_{bc}$, so that bosons are in a Mott state in the absence of 
Feshbach resonances.
On the other hand $r_f$ is much 
bigger than the critical value $r_{fc}$ so 
that Cooper pairs are condensed.
For simplicity, we have also assumed that $V_{bf}$ is much smaller than $V_b$ so that
it can be treated as a perturbation.
We are interested in the responses of bosonic Mott states to 
fermion-boson conversion  
and carry out the rest of discussions in a mean field approximation
({\em MFA}).

\begin{figure}
\centering\epsfig{file=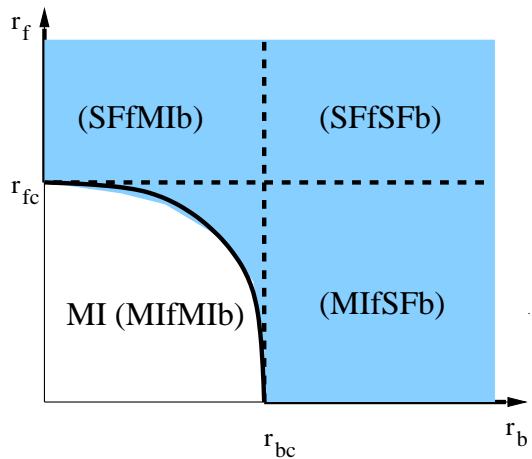,clip=1,width=0.8\linewidth,angle=0}
\caption{ Phase diagrams with (solid lines) and without (Dashed lines) 
Feshbach fermion-boson conversion.
$r_{fc}$ and $r_{bc}$ are the critical values for the superfluid-Mott
insulator transitions for the decoupled fermionic Cooper pairs and 
bosonic molecules respectively.
Phases in brackets are the ones 
without Feshbach fermion-boson conversion and are separated by dashed lines.
Note in the presence of fermion-boson conversion,
due to the invasion of superfluidity to MI phases,
the original SFfSFb, SFfMIb, MIfSFb phases, and a small portion of the MIfMIb phase
merge into one single superfluid phase specified as the shaded area.
}
\end{figure}

In this {\em MFA}, $\phi_{fk}=\phi_f$, $\phi_{bk}=\phi_b$ 
for any lattice site $k$. The ground state
$\Psi_0(\phi_b, \phi_f)$ (again defined on an $S^1\otimes S^1$ torus with radius $2\pi$) is 
the lowest energy state of the 
following {\em MFA} Hamiltonian
\begin{eqnarray}
H_{MFA} &=&- V_f \frac{\partial^2}{\partial \phi^2_{f}}
-V_b \frac{\partial^2}{\partial \phi^2_{b}}
-2V_{bf} \frac{\partial}{\partial \phi_{f}}\frac{\partial}{\partial \phi_{b}}
\nonumber \\
&-& z (J_f  \Delta_f  \cos\phi_{f} + J_b  \Delta_b \cos\phi_{b} ) 
\nonumber \\
&-& \Gamma_{FB} \cos (\phi_{f}-\phi_{b}).
\label{MFA}
\end{eqnarray}
Here again $z$ is the coordination number; we have also introduced two 
self-consistent order parameters
\bea
&& \Delta_{b,f}= \langle\cos\phi_{b,f}\rangle=
\int_0^{2\pi} {d\phi_f} \int_0^{2\pi} {d\phi_b}
\cos\phi_{b,f} \Psi_0\Psi_0^*. \nonumber \\
\label{OP0}
\eea
Here $\langle\rangle$ stands for an average taken in the ground state,
and $\Psi_0$ is the ground state wavefunction.
Notice that the order parameters defined above are nonzero only when the 
U(1) symmetries are 
broken; particularly, $\Delta_f$ is proportional to the usual BCS pairing 
amplitude. Following Eq.(\ref{coherent}) and discussions above one indeed 
shows that
\bea
&& \tilde{\Delta}_f=<f^\dagger_{k\eta\uparrow}f^\dagger_{k\eta\downarrow}> =(\sum_{\eta} u_\eta 
v_\eta) \Delta_f,
\nonumber \\
&& \tilde{\Delta}_b= <b^\dagger_{k}>=\sqrt{n_b} \Delta_b.
\eea
So $\Delta_{f,b}$ vanish in Mott states and are nonzero 
in superfluids.

As $z J_f$ is much larger than $V_f$, 
$\phi_f$ has very slow dynamics; and the corresponding ground state for $\phi_f$ 
can be approximated as a symmetry breaking solution.
In the linear order of $J_b$ and $\Gamma$, one obtains the following 
solution
\begin{eqnarray}
&& \Psi_0(\phi_f, \phi_b)=\Psi_{0b}(\phi_b) \otimes \delta(\phi_f),
\nonumber \\
&& \Psi_{0b}(\phi_b)=\frac{1}{\sqrt{2\pi}}[1
+(\frac{z J_b}{V_b} \Delta_b 
+\frac{\Gamma_{FB}}{V_b}\Delta_f) \cos\phi_b]. \nonumber 
\\
\label{wf}
\end{eqnarray}
Here $\Delta_f$ should be approximately equal to one in this limit;
and in the zeroth order of $V_b^{-1}$, $\Psi_{0b}$ does not break the 
$U(1)$ symmetry and stands for a Mott-state solution.
Finally taking into account Eq.(\ref{OP0}) and (\ref{wf}),
one finds that the self-consistent solution 
to $\Delta_b$ is
\begin{equation}
\Delta_b= \frac{1}{2}\frac{\Gamma_{FB}}{V_b}
[1-\frac{1}{2}\frac{z J_b}{V_b}]^{-1}.
\label{order}
\end{equation}
In the absence of $\Gamma_{FB}$, $\Delta_b$ vanishes as expected for a Mott state.
However, the Mott state solution is unstable in the presence of any Feshbach 
conversion and the molecular condensation order parameter
$\Delta_b$  is always nonzero in this limit.

We want to emphasize that the {\em average} number of bosons per site is not 
affected by the fermion-boson conversion and remains to be an integer 
($I$); rather, 
closely connected with the instability is the breakdown of particle number quantization.
Indeed, one obtains in the {\em MFA} the following results for 
$\hat{n}_{bk}$,
\begin{equation}
<\hat{n}_{bk}>=n_b=I, <\delta^2 
\hat{n}_{bk}>=1/2 ({\Gamma_{FB}}/{V_b})^2 \approx 2 \Delta_b^2.
\end{equation}
This illustrates that the resonance between states with different numbers of 
bosons at a lattice site eventually leads to a nonzero 
molecular condensation order parameter $\Delta_b$.

Alternatively, one can consider the renormalization of the condensate 
amplitude
due to enhanced quantum fluctuations when repulsive interactions are 
introduced. When repulsive interactions are weak, one can carried out 
usual perturbative calculations. 
\begin{figure}
\centering\epsfig{file=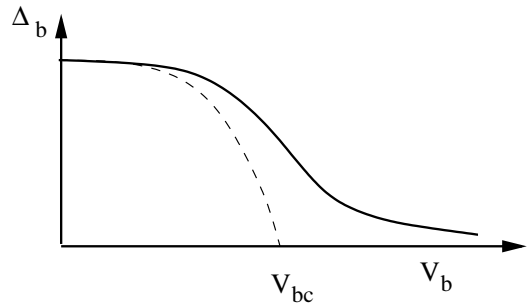,clip=1,width=0.8\linewidth,angle=0}
\caption{ Schematic of the renormalized condensate amplitude $\Delta_b$
as a function of $V_b$ ($t_b$ is given and set to be unity).
The solid line and dashed line are for the case 
with and without fermion-boson conversion respectively. }
\end{figure}

The results above on the other hand provide information about
what happens when interactions are dominating and the conventional 
perturbation 
expansion fails. One of the most important 
consequences of fermion-boson conversion is that the suppression of condensate 
amplitude is never complete if 
one increases $V_b$ only while maintaining small
value of $V_f$. The renormalization of the condensate amplitude as 
a function of $V_b$ is plotted schematically in Fig.4.

At last, let us briefly consider the case that both bosons and fermion pairs
are in Mott states, i.e., $r_{b}<r_{bc}$ and $r_{f}<r_{fc}$ respectively.
Then the mass gaps in two channels behave like
$m_b\propto r_{bc}-r_{b}$ and $m_c\propto r_{fc}-r_f$
in the absence of the boson-fermion conversion term $\Gamma_{FB}$.
The conversion term leads to a hybridization between two components,
and we can diagonalize their mass matrices to find the new
mass gaps.
When $\Gamma^2>m_f m_b\propto (r_{bc}-r_b) (r_{fc}-r_c)$, 
one eigenvalue becomes negative which
suggests that the Mott state should be unstable, i.e.,
a superfluid state should be formed.

\subsection{Saddle point approximation}
In this subsection, we are going to provide an alternative 
approach to ODLO based on a saddle point approximation. 
We will show again that any finite fermion-boson
conversion leads to a finite condensation of bosons disregarding 
the strength of repulsive interactions between bosons.

For this purpose we first introduce the following partition function
\bea
Z&=&\int D f^\dagger  Df D b^\dagger D b ~~~ \exp\{ -(
S_f+ S_b + S_{bf} ) \} 
\label{eq:action}
\\
S_f&=& \int^\beta_0 d\tau \int d\vec r \ \ \  f^\dagger_\sigma(\vec r)
\Big \{ \frac{\partial}{\partial\tau} +\epsilon(\nabla)-\mu) 
\Big \}
f_\sigma(\vec r) \label{eq:fermibcs}\\
&-& g f^\dagger_\uparrow(\vec r) f^\dagger_\downarrow (\vec r)
f_\downarrow(\vec r) f_\uparrow(\vec r)   \\
S_b&=& \int^\beta_0 d\tau   \ \ \
\sum_k  b^\dagger_k  \frac{\partial}{\partial \tau} b_k
 -t\sum_{\avg{kl}} \Big\{ b^\dagger_k b_l +h.c. \Big\}\nonumber  \\
&-&2 \mu \sum_k b_k^\dagger b_k
+\frac{U}{2}\sum_k b^\dagger_k b_k (b^\dagger_k b_k-1)
\label{eq:bsHub}
\\
S_{bf}&=& \int^\beta_0 d\tau \sum_k  \ \ \
\Gamma \Big\{ b^\dagger_k  f_\downarrow(k) f_\uparrow(k)
+b_k f^\dagger_\uparrow(k) f^\dagger_\downarrow(k)
\Big\} \label{eq:bsfermi}, \nonumber \\
\eea
where $f_\sigma(\vec r),f_\sigma^\dagger (\vec r)$ are fermion field variables
defined in the continuum,
$b_k$ are the field variables for bosonic molecules in the closed channel
defined at lattice site $k$,
$f_\sigma(k)$ is a coarse average of $f_\sigma(\vec r)$ within 
the $k$-th  unit cell as
\bea
f_\sigma(k)=\frac{1}{\Omega} \int_{\Omega_k} d \vec r  \ \ \ f_\sigma(\vec r),
\eea
where $\Omega$ is the volume of one unit cell.

For the reasons outlined in Appendix A, it is more convenient to introduce
a $\phi$-field variable to describe the dynamics of bosons; the $\phi$-field
can be interpreted as the condensate wavefunction.
Following discussions
there, in the long wave length limit we obtain the following $\phi^4$-theory description of bosons
\bea
S_\phi&=& \int^\beta_0 d\tau \int d\vec r  \ \ \
\phi^\dagger(\vec r) \Big\{ 
r^\prime \frac{\partial}{\partial\tau}+
r \frac{\partial^2}{\partial \tau^2}
-\kappa \nabla^2+\alpha \Big \}  \nonumber \\
&\times& \phi(\vec r)+\frac{\lambda}{2} (\phi^\dagger\phi)^2
\label{eq:sf}
\\
S^\prime_{bf}&=& \int^\beta_0 d\tau \int d\vec r  \ \ \
\Gamma \Big\{ \phi^\dagger(\vec r) f_\downarrow(\vec r) f_\uparrow(\vec r)
\nonumber\\
&+&\phi(\vec r) f^\dagger_\uparrow(\vec r) f_\downarrow(\vec r)
\Big\}.
\eea
According to Appendix A, when the system has a particle-hole symmetry,
one can set $r'$ to be zero. This also corresponds to a system where $n_b$ is an integer,
a situation we have discussed in the previous session.
Moreover, when $\alpha>0$$( <0)$, the system is in a Mott phase with an energy gap
(in the superfluid phase). 

Using the standard Hubbard-Stratonovich transformation, we decouple the
4-fermion interaction term by introducing the pairing field $\Delta$.
After integrating out the fermions in $S'_{bf}$, $S_f$ (see Appendix A, B and especially 
Eq. \ref{eq:HS}, \ref{eq:integral}), we arrive at
the following effective action
\bea
S_{eff.} &=& S_\phi + S_{\Delta\phi}\nonumber \\
S_{\Delta\phi} &=&\det \Big \{ 
\frac{\partial}{\partial \tau}+ 
G_1\tau_1+G_2 \tau_2+ G_3 \tau_3
\Big \},
\nonumber \\
G_1 &=&
-(\Re \Delta(\vec r,\tau ) +
\Gamma ~\Re \phi (\vec r, \tau)), 
\nonumber \\
G_2 &=&
-(\Im \Delta(\vec r,\tau ) + \Gamma ~\Im \phi (\vec r, \tau)), 
\nonumber \\
G_3 &=&\epsilon(\nabla)-\mu),
\label{effective}
\eea
where $\tau_{1,2,3}$ are the Pauli's matrices in the Nambu's representation,
and $\tau_{+,-}=(\tau_1\pm i\tau_2)/2$.

Consider the standard mean field ansatz
\bea
\Delta(\vec r,\tau)= \bar \Delta +\delta \Delta \ \ \
\phi(\vec r, \tau)= \bar \phi +\delta \phi,
\eea
Taking into account the contribution from molecules in Eq. \ref{eq:sf},
we then obtain the standard self-consistent equation for $\bar \Delta$
and $\bar \phi$
\bea\label{eq:saddle}
&&\frac{\bar \Delta}{g}-\frac{1}{V} \sum_{\vec k}
\frac{\bar \Delta +\Gamma \bar \phi} {2 E_k} =0,
\nonumber \\
&& \alpha \bar \phi+\lambda |\bar \phi|^2 \bar \phi
-\frac{\Gamma}{V} \sum_{\vec k}
\frac{\bar \Delta +\Gamma \bar \phi} {2 E_k} =0.
\label{eq:saddle1}
\eea
The dispersion relation for quasi-particles is
\bea
E_k^2=(\epsilon_k-\mu)^2 +|\bar \Delta +\Gamma \bar \phi|^2.
\eea

Eq.(\ref{eq:saddle1}) is of the same form as the equation for $\Delta,\bar{\phi}$ derived in a zero-V limit
where bosons and cooper pairs are non-interacting.
However, in our case, the equation is valid even when various repulsive interactions are strong and bosons are in a
Mott state.
We derive the relation between $\bar\phi$ and $\bar\Delta$
from Eq. \ref{eq:saddle1} as
\bea
\alpha \bar \phi +\lambda |\bar \phi|^2 \bar \phi
= \Gamma \frac{\bar \Delta}{g}.
\label{eq:relation}
\eea
It is clear the saddle point values of $\bar \phi$ and $\bar \Delta$
are locked with the same phase. For convenience, we assume both of
them to be real in the following.

First consider a situation when molecules are deeply in the Mott state, i.e. $\alpha\gg \Gamma,
\Delta>0$. Eq. \ref{eq:relation} can be approximated up to
$\Gamma$'s second order as
\bea
\frac{\bar \phi}{\bar \Delta/g}
=\frac{\Gamma}{\alpha}.
\eea
This equations shows that condensation amplitude of bosons is finite for
any finite coupling $\Gamma$,
disregarding the value of $\alpha$.
When $\Gamma$ is nonzero, this equation indicates that the minimum for the total energy should be located at
a finite $\bar{\phi}$ instead of zero.

On the other hand, if molecules are deeply in the superfluid state, i.e.,
$-\alpha \gg \Gamma>0$, the solution to Eq. \ref{eq:relation} can be approximated as
\bea
&& \bar \phi= \phi_0 + \phi^\prime,
\eea
where $\phi_0$ is the saddle point value without the boson-fermion
conversion, and $\phi^\prime$ is the correction,
\bea
&&\phi_0=\sqrt{\frac{|\alpha|}{\lambda}  }, \ \ \  \ \ \
\phi^\prime=-\frac{1}{2}\frac{\Gamma}
{\alpha} \frac{\bar \Delta}{g}.
\label{eq:moleculesf}
\eea
It is worth remarking again
that in this limit the phase of $\bar{\phi}$ is precisely locked with 
the phase of $\bar \Delta$, following Eq.(\ref{eq:relation}).
This is consistent with the Hamiltonian-based discussion 
in section IV A.

\section{ODLO in the Single Band limit}
The main conclusions arrived so far do not depend on the {\em large-N} 
approximation introduced above.
 One can consider the opposite limit by assuming $M=1$ and there 
is only one orbital degree of freedom at each lattice site.
In the single-orbit 
limit,  the two interaction terms 
(with two interaction constants $\lambda$ and $V_f$ ) in $H_f$
(see Eq. (\ref{FBHM})) 
can be rewritten in one term: 
$V_f' \sum_k \hat{n}_{fk}(\hat{n}_{fk}-1)$ if one identifies
$V_f'=V_f -2 \lambda$.
In the limit where $\lambda$ is much larger than $V_f$, fermions are paired 
at each lattice site. Furthermore, I assume bosons have hard core 
interactions ($V_b=\infty$) such that there can be only zero or one boson at each site.

So the low energy Hilbert subspace ${\cal S}_k$ at each lattice site $k$ 
consists of four states: 1) no Cooper pair, no boson; 2) no Cooper pair, one boson;
3) one Cooper pair, no boson; 4) one Cooper pair, one boson.
They also correspond to a product of two pseudo spin $S=1/2$ 
subspaces: 
\bea
&& {\cal S}_k={\cal S}_{fk}\otimes {\cal S}_{bk},\nonumber \\
&& |\sigma^{z}_{fk}=\pm 1 \rangle \in {\cal S}_{fk},
|\sigma^{z}_{bk}=\pm 1 \rangle \in {\cal S}_{bk};
\eea
${\cal S}_k$ is the on-site Hilbert space, and
${\cal S}_{fk,bk}$ are the on-site pseudo spin spaces 
for fermions and bosons respectively.
More explicitly, these four states are
\begin{eqnarray}
&& |\sigma^z_{fk}=1\rangle={f^\dagger_{k\uparrow}f^\dagger_{k\downarrow}}
|vac\rangle_f,\nonumber\\
&& |\sigma^z_{bk}=-1\rangle=|vac\rangle_f; \nonumber \\ 
&& |\sigma^z_{bk}=1\rangle=b^\dagger_k |vac\rangle_b,\nonumber \\
&&|\sigma^z_{bk}=-1\rangle=|vac\rangle_b.
\end{eqnarray}
$|vac\rangle_{f,b}$ are the vacuum of fermion and bosons respectively.
Finally, in this truncated subspace, the following identities hold
\begin{eqnarray}
&& \sigma^+_{fk}=f^\dagger_{k\uparrow}f^\dagger_{k\downarrow},
\sigma^-_{fk}=f_{k\downarrow}f_{k\uparrow},\nonumber\\
&& \sigma^z_{fk}=
f^\dagger_{k\uparrow}
f_{k\uparrow}
+f^\dagger_{k\downarrow}
f_{k\downarrow}
-1; \nonumber\\
&& \sigma^+_{bk}=b^\dagger_{k},
\sigma^-_{bk}=b_{k}, \sigma^z_{bk}=2b^\dagger_{k} b_k-1.
\end{eqnarray}
So to have superfluidity, either $\sigma_{bk}$
or $\sigma_{fk}$, or both of them need to have a finite expectation value in the {\em XY} plane. 
For instance to have fermionic superfluids, the expectation values 
of $\sigma^\pm_{fk}$ need to be nonzero.

The effective Hamiltonian can then be written as

\begin{eqnarray}
&& H^1_{eff}=-J^1_b \sum_{\langle kl\rangle}
\Big\{  \sigma^x_{bk} \sigma^x_{bl} + \sigma^y_{bk}\sigma^y_{bl} \Big\}
-h^z_b \sum_k \sigma^z_{fb} \nonumber \\
&& 
-J^1_f \sum_{\langle kl \rangle} 
\Big\{\sigma^x_{fk} \sigma^x_{fl} + \sigma^y_{fk}\sigma^y_{fl}
-\sigma^z_{fk}\sigma^z_{fl} \Big\}
-h^z_f \sum_k \sigma^z_{fk}  \nonumber \\
&&-\Gamma^1_{FB} \sum_{k} \Big\{ \sigma^x_{fk} \sigma^x_{bk} 
+ \sigma^y_{fk}\sigma^y_{bk} \Big\}.
\end{eqnarray}
(see also Ref.\onlinecite{Zhou05})
Here 
\bea
&& J_f^1=t_f^2/V'_f, J^1_b=t_b, 
\nonumber \\
&& \Gamma^1_{FB}=\gamma_{FB}, h^z_f= \mu + 
V'_f/2, h^z_b=\mu - v/2.
\eea
The Hamiltonian is invariant under a rotation around the $z$-axis
or has an $XY$ symmetry.
The z-direction {\em fully} polarized phase of pseudo spins $\sigma_{bk}$ 
($\sigma_{fk}$) represents the Mott phase of bosons (fermions),
and the $XY$ symmetry breaking 
states of pseudo spins $\sigma_{bk}$ ($\sigma_{fk}$) stand for the superfluid phase of bosons 
(fermions). The fermionic sector of this Hamiltonian was previously obtained and 
studied\cite{Emery76}; it was also used to study BEC-BCS crossover in lattices\cite{Carr05}.

When $\Gamma^1_{FB}=0$, the Mott phase for bosons with filling factor 
equal to 
one occurs when $h_{b}^z$ is much larger than $J_b$. Assume that in this case $h_f^z$ 
is much less than 
$J_f$ so that $\vec{\sigma}_{fk}$ are ordered in the $xy$ plane; 
then fermions form Cooper pairs. 
Taking into account a finite amplitude of $\Gamma^1_{FB}$, one 
considers a solution where the pseudo spin symmetry of $\vec{\sigma}_{fk}$
is spontaneously broken along a direction in the XY plane specified by 
$\avg{\vec{\sigma}_{fk}}$
(the expectation value is taken in the ground state). 

In the molecular mean field  approximation, the effective
external field acting on pseudo spins ${\vec \sigma}_{bk}$ is
\begin{eqnarray}
{\vec h}_{b, eff}= z J_b \langle \vec {\sigma}_{bk} \rangle +  \Gamma^1_{FB} 
\avg{{\vec \sigma}_{fk}}  + h^z_b {\vec e}_z.
\end{eqnarray} 
$\avg{{\vec {\sigma}}_{bk}}$ 
is calculated self-consistently in the ground 
state when ${\vec h}_{b, eff}$ is applied;
the effective MFA Hamiltonian is
\bea
H_{MFA}=-\vec \sigma_{bk} \cdot {\vec h}_{b,eff}.
\eea
One then arrives at the following self-consistent solution 
\begin{eqnarray}
\avg{{\vec \sigma}_{bk}} \cdot  \avg{{\vec 
\sigma}_{fk}}\approx \frac{\Gamma^1_{FB}}{h^z_b}(1-\frac{zJ_b}{h_b^z})^{-1},
\end{eqnarray}
where $\avg{{\vec \sigma}_{bk}}$ has been projected along the 
direction of $\avg{{\vec \sigma}_{fk}}$ which lies in the $XY$ plane. As 
mentioned before,
development of such a component signifies superfluidity, or 
molecular condensation.

To summarize, we have shown that certain Mott states are unstable with 
respect 
to the resonating fermion-boson conversion;
in general superfluidity invades Mott phases because of the fermion-boson 
conversion. 

\section{Hidden order and vortices in Mott states}
In addition to introducing superfluidity to Mott states in some limit, the fermion-Boson
conversion also results in a hidden order in Mott states. 
In the presence 
of fermion-boson conversion, one finds it is more convenient to introduce
trilinear order parameters to characterize a Mott state
\begin{equation}
\Delta^+_{bf}=\langle
b^\dagger_k f^\dagger_{k\eta\sigma}f^\dagger_{k\eta-\sigma}
\rangle, \ \ \
\Delta^-_{bf}=\langle b^\dagger_k f_{k\eta\sigma}f_{k\eta-\sigma}\rangle.
\end{equation}
A superfluid phase would have a nonzero order parameter of
$\Delta^+_{bf}$ type, but $\Delta^-_{bf}$ can be either zero or nonzero.
For a usual superfluid near Feshbach resonances, $\Delta^-_{bf}$ is nonzero. 
However, there might be
more exotic superfluids where $\Delta^-_{bf}$ is zero; when this occurs, 
the superfluid
will have two decoupled components with {\em unlocked} phases.
On the other hand, a Mott state has vanishing $\Delta^+_{bf}$ but 
always has non-vanishing $\Delta^-_{bf}$ 
as hidden order as far as the fermion-boson conversion 
is present(see below). 
Here the appearance of $\Delta^-_{bf}$ order is due to
the fermion-boson conversion. 
Effectively, it can be viewed as an order parameter of
a boson and a Cooper-pair hole pairing, which bears resemblance of the 
electron-hole exciton formation in semiconductors \cite{Mahan00}
and in quantum Hall bilayer systems \cite{Eisenstein04}.

To understand this issue, 
we first consider an extreme situation when 
$V_f=V_b=V_{bf}=V_0$ and all of them
are much larger than $t_{b,f}$ and $\gamma_{FB}$. Minimizing the potential energy leads
to the following constraint on $n_{fk,bk}$ in the ground state
\begin{equation}
N_{fb}=\frac{n_{fk}}{2}+n_{bk}=\mbox{Int}[\frac{3\mu-v }{V_0}+\frac{5}{4}].
\end{equation}
Here $I$ is an integer, $\mbox{Int}[I+\epsilon]$ is equal to $I$ if $0\leq 
\epsilon <1/2 $,
and to $I+1$ if
$1/2 < \epsilon \leq 1$; at $\epsilon=1/2$, Int takes either $I$ or $I+1$.
Let us assume that the chemical potentials and interactions are such that 
$N_{fb}$ is equal to an integer $I$. 
All states satisfy the constraint are degenerate when $\gamma_{FR}$ is zero,
and thus the degeneracy is proportional to
$N_{fb}$.

In this limit, we can truncate the Hilbert space and consider the effect of Fermion-boson conversion in the degenerate 
subspace only.
We study the following  ground state trial wavefunctions
constructed out of these degenerate states,
\begin{eqnarray}
|g\rangle&=&\prod_k  
\exp\{-i\phi_{kfb}n_{kfb}\}
\sum_{n_{kfb}<N_{fb}}
\Big\{
\frac{(b^\dagger_k)^{N_{fb}-n_{kfb}}}{\sqrt{(N_{fb}-n_{kfb})!}}
\nonumber\\
&\times&
\sum_{\{n_{k\eta}\}}
 \prod_{\eta} w(n_{{k\eta}})(f^\dagger_{k\eta\uparrow} 
f^\dagger_{k\bar\eta\downarrow})^{n_{k\eta}} \Big\}
|0\rangle,
\label{trial}
\end{eqnarray}
where $n_{k\eta}=0$ or $1$ satisfying $\sum_\eta n_{k\eta} =n_{kfb}$,
$w(n_{k\eta})=u_\eta$ at $n_{k\eta}=0$ and $v_\eta$ at $n_{k\eta}=1$,
respectively ($u_\eta, v_\eta$ are coherence factors).
One can easily verify that states with different $\{\phi_{kfb}\}$ are 
approximately orthogonal when $N_{fb}$ is much larger than unity.

But any finite conversion leads to a lift of degeneracy.   
The energy associated with the conversion is
\begin{equation}
E\sim -\Gamma_{FB}\sum_k \cos(\phi_{kfb}).
\end{equation}
Minimization takes place when $\phi_{kfb}=0$ for any lattice site $k$.
The symmetry here is broken not spontaneously as in superfluids 
but actually broken explicitly by the fermion-Boson conversion.
The ground state is non-degenerate and 
does not have the usual
$U(1)$ vacuum manifold.

This state is characterized by the following expectation values
\begin{eqnarray}
&& \avg{b^\dagger_k}=\avg{f^\dagger_{k\eta\sigma}f^\dagger_{k\eta-\sigma}}
=\avg{b^\dagger_kf^\dagger_{k\eta\sigma}f^\dagger_{k\eta-\sigma}}=0
\nonumber \\
&& \avg{b^\dagger_kf_{k\eta\sigma}f_{k\eta-\sigma}}\sim N_{fb}.
\end{eqnarray}
The existence of the trilinear order in Mott states is very unique
and defines a hidden order.
There are a few consequences.
One is the collective excitations.
In addition to excitations which have an energy gap $V_0$,
there are another branch of excitations involved 
the creation of a bosonic
particle and annihilation of a cooper pair,
$\sum_{q_1, q_2}
b^\dagger_{q_1+Q} f_{q_2\eta\uparrow}f_{-q_2+q_1\eta\downarrow}$;
these excitations are gapped by the energy of order $\Gamma_{FB}$
instead of the Mott gap.

Furthermore, a hidden order also implies new classes of topological 
excitations.
The wavefunction of a topological excitation centered at 
the origin is given by Eq.(\ref{trial}) where
$\phi_{kfb}$ is defined by the following equation
\begin{equation}
\phi_{kfb}=\Phi(R_k);
\end{equation}
$\Phi(R_k)$ is the azimuthal angle of $R_k$. 
The vortex is orientated along the $z$-direction.
The energy per unit length of this excitation unfortunately scales as the
area of the system in the $xy$ plane; i.e. 
\begin{equation}
\frac{E_v}{L_z}\sim \Gamma_{FR} \sum_k [ 1-\cos(\phi_{kfb})]=L_xL_y 
\Gamma_{FB}.
\end{equation}

The situation discussed here is not generic and requires fine tuning.
Let us now turn to a more general situation where $V_{b}\neq V_f$.
If $r_{b,f}$ are much smaller than $r_{bc,fc}$, then both Cooper pairs and
bosons are in Mott states. 
Up to the first order approximation of $\Gamma_{FB}$,
the corresponding wavefunction is

\begin{eqnarray}
&&|g\rangle \approx \prod_k  
\Big\{
\frac{(b^\dagger_k)^{n_b}}{\sqrt{n_b!}}
\sum_{\{n_{k\eta}\}}
 \prod_{\eta} w(n_{k\eta})(f^\dagger_{k\eta\uparrow} 
f^\dagger_{k\bar\eta\downarrow})^{n_{k\eta}} \nonumber \\
&+& e^{\pm i\phi_{kfb}} \frac{\Gamma_{FB}}{V_b+V_f} 
\frac{(b^\dagger_k)^{n_b\pm1}}{\sqrt{(n_{fb}\pm1)!}}
\nonumber \\
&\times&\sum_{\{n^\prime_{k\eta}\}}
 \prod_{\eta} w(n^\prime_{k\eta})(f^\dagger_{k\eta\uparrow} 
f^\dagger_{k\bar\eta\downarrow})^{n^\prime_{k\eta}}\Big\}
|0\rangle,
\label{trial2}
\end{eqnarray}
where the distribution $n_{k\eta}^\prime$ satisfies
$\sum_\eta n_{k\eta}^\prime=n_f/2\pm1$, and $n_{k\eta}$ satisfies
$\sum_\eta n_{k\eta}=n_f/2$. 
$\phi_{kfb}$ has to be uniform and zero for the ground state.

Similar calculations lead to
self-consistent solutions $\Delta_f=\Delta_b=0$ and more importantly, the following correlations
for $\Delta_{bf}^\pm$,

\begin{eqnarray}
\Delta_{bf}^-\approx \frac{\Gamma^2_{FB}}{2\gamma_{FB}(V_f+ V_b)}, 
\Delta^+_{bf}=0.
\end{eqnarray}
The second equality above simply shows the absence of superfluidity.
But the first one indicates a subtle {\em hidden} order in the Mott states under 
consideration. Notice that $\Delta_{bf}^{\pm}$ represent 
tri-linear order and are proportional to 
$\avg{\cos(\phi_{kb}\pm \phi_{kf})}$.

One can easily show that the vortex wavefunction is given by the same 
expression but with $\phi_{kfb}=\Phi(R_k)$; the energy per unit length in 
this case is much smaller
\begin{equation}
\frac{E_V}{L_z}\sim L_xL_y \frac{\Gamma^2_{FB}}{V_f+V_b}.
\end{equation}

\section{Conclusions and Acknowledgment}
In this article, we examine the stability of bosonic Mott states 
under the influence of fermion-boson
conversion and study various aspects of Mott states and superfluids 
when repulsive interactions among 
bosons are very strong. We have found that when 
bosonic Mott states are coupled to fermionic superfluids
via the fermion-boson conversion, 
there appears to be a finite condensation fraction of bosons in the ground 
state. There are 
extra low energy states below the Mott gap representing gapless {\em 
extended} excitations in 
the Mott-superfluid
mixture, due to delocalization of bosons. We also show the existence of 
off-diagonal long-range order in the 
bosonic channel due to fermion-boson conversion.

The second issue we look into here is the novel collective excitations 
in superfluids.
This branch of excitations involves oscillations
of the difference between the boson and fermion density. 
Unlike the usual Goldstone modes in superfluids, it is fully gapped. 
The gap energy is proportional to $\Gamma_{FB}$ when bosons are 
weakly interacting, or in a superfluid state.

At last, we study the Mott states of boson-fermion mixture and 
find hidden trilinear order in Mott states. We analyze the order
in a few limits and briefly study the novel topological excitations
in Mott states.

FZ is in part supported by a Discovery grant from NSERC, Canada and 
a research grant from the Dean of Science's office, UBC; FZ is also an A. 
P. Sloan fellow. CW is supported by the NSF Grant No. Phy99-07949. 
FZ wants to  thank CASTU, 
Tsinghua University, Beijing and INT, University of Washington, Seattle 
for hospitalities.
Finally, both authors want to thank ASPEN center for physics for its 
hospitality during the workshop "Quantum gases 2005".

\appendix

\section{The Effective Action}
In this section, we present the effective action for the boson-fermion
conversion in the optical lattices near Feshbach resonances.
We start with the microscopic actions of Eq. \ref{eq:action},
\ref{eq:fermibcs}, \ref{eq:bsHub}, \ref{eq:bsfermi} in Sec. VB.
Eq. \ref{eq:fermibcs} describes the attraction 
in the open channel for the formation of BCS Cooper pairs;
Eq. \ref{eq:bsHub} describes the Bose-Hubbard (BH) model of the boson molecules
with $t$ the hopping integral and $U$ the on-site repulsion;
Eq. \ref{eq:bsfermi} describes the conversion between the
Cooper pairs and molecules.

The BH model of Eq. \ref{eq:bsHub} exhibits a superfluid 
(SF)-Mott insulating (MI) phase transition \cite{Fisher89}. The MI
phase only exists in the strong coupling regime with commensurate fillings,
i.e., small values of $t/U$ and integer values of $n_b$.
The SF-MI transition can be obtained by two different ways.
First, the boson filling is kept commensurate while $t/U$ is tuned
larger than the corresponding critical value.
This transition belongs to the XY universal class, and the resulting
SF is particle-hole symmetric.
Second, we can also add or remove particles to the commensurate MI
background, i.e., dope the MI with extra particles or holes.
Because the particle-hole symmetry is broken, this transition is not
XY-like.
The resulting SF are either particle-like or hole-like.
Consequently, although only one connected SF phase exists in the phase 
diagram, it actually exhibits rich
structures, including the particle-like, hole-like,
or even relativistic (particle-hole symmetric) SF,
which are connected by smooth cross-overs.

The bare boson operators $b_k,b_k^\dagger$ in Eq. \ref{eq:bsHub} 
are for non-relativistic particles.
However, near the SF-MI transition,
it is not convenient to use them
to describe above rich structures in the SF phase.
For example, $b_k$ means both an annihilation of
a particle and a creation of a hole in the MI background.
On the other hand, the SF-MI transition is in the strong
coupling regime by using the bare operators of $b_k,b_k^\dagger$, and
it is hard to do perturbation theory for the Hubbard $U$ term.
Thus we follow Fisher et al.  \cite{Fisher89} to introduce
another complex bose field $\phi (\vec r)$ to describe the molecular
superfluidity. This can be formally done by keeping the on-site
Hubbard term in  Eq. \ref{eq:bsHub} as the leading term, 
and decoupling the inter-site hopping term as perturbations.
Basically, this transformation turns the original strongly interacting
non-relativistic systems into  weakly interacting 
quasi-relativistic  systems.
It is shown in the equation of motion that
$\phi$ plays the role of the expectation value of $b$
in the ground state, i.e.,
$\phi$ is the superfluid order parameter.

Phenomenologically, the symmetry allows an effective action 
for the $\phi$-field upto the quartic level as
\bea
S_\phi&=& \int^\beta_0 d\tau \int d\vec r  \ \ \
\phi^\dagger(\vec r) \Big\{ 
r^\prime \frac{\partial}{\partial\tau}+
r \frac{\partial^2}{\partial \tau^2}
-\kappa \nabla^2+\alpha \Big \}  \phi(\vec r) \nonumber \\
&+& \frac{\lambda}{2} (\phi^\dagger\phi)^2,
\label{eq:sf1}
\eea
which includes the $r,\kappa,\alpha,\lambda$ terms as in the standard
relativistic complex $\phi^4$ theory, 
and also an additional first order time derivative term of 
$r^\prime$.
Whether the mass $\alpha>0$ or $ \alpha<0$ determines the system either 
in the MI phase with a charge gap or in the SF phase, respectively.
All these coefficients in Eq. \ref{eq:sf1} 
can be  determined by the values of $t, U, \mu$ 
in the original BH model perturbatively \cite{Fisher89}.
However, for simplicity we treat them as phenomenological parameters.
It is proved through gauge invariance 
that $r^\prime$ is related with $\alpha$ through \cite{Fisher89}
\bea
r^\prime = -\frac{\partial \alpha}{\partial \mu}.
\eea
Near the SF-MI transition, 
as the filling $n_b$ changes from one integer to another
integer, the superfluidity is enhanced and suppressed alternatively.
As a result, $\alpha$ oscillates, and then $r^\prime$ can be
negative, positive, or even zero.
Roughly speaking, when $n$ is larger (smaller) than an integer number,
$r^\prime>0$ ($r^\prime<0)$, and then the system is particle-like (hole-like).
As long as $r^\prime\neq 0$, the first order time derivative term
dominates over the second order one below a certain energy scale
in the sense of the renormalization group (RG),
and the system is non-relativistic.
When $n_b$ is commensurate,  the superfluidity is in a local
minimum, and thus $r^\prime=0$, i.e., the system is particle-hole symmetric.
In other words,  the $r$ term becomes the leading order term, and the system
becomes  relativistic. 

Many possible terms coupling the superfluid field $\phi$ 
and fermions $f_k^\dagger,f_k$ together are  allowed by symmetry.
Among them,  the linear coupling term is the most relevant one 
in the sense of RG as
\bea
S^\prime_{bf}&=& \int^\beta_0 d\tau \int d\vec r  \ \ \
\Gamma \Big\{ \phi^\dagger(\vec r) f_\downarrow(\vec r) f_\uparrow(\vec r)
+c.c.
\label{eq:bsfermicnt}
\Big\}.
\eea
Here we use the same symbol $\Gamma$ for the coupling constant
as in Eq. \ref{eq:bsfermi} for convenience.
However, we need to bear in mind that the  
$\Gamma$ here receives significant renormalization
from its bare value in Eq. \ref{eq:bsfermi}.

The action for the fermion BCS interaction in Eq. \ref{eq:fermibcs} is
already defined in the continuum. Combined with Eq. \ref{eq:sf1} 
and Eq. \ref{eq:bsfermicnt}, these three terms give 
the effective action for the boson-fermion conversion.

\section{Self-consistent equation}
In this section, we derive the self-consistent equations for
the coupled superfluids of bosonic molecules and the fermionic Cooper pairs.
Using the standard Hubbard-Stratonovich (HS) transformation,
we decouple the 4-fermion interaction term in Eq. \ref{eq:fermibcs}
in terms of Cooper pair filed $\Delta$ as
\bea
&&\int D f D f^\dagger \exp \Big \{\int^\beta_0 d\tau \int d\vec r 
\Big \{ g f^\dagger_\uparrow(\vec r) f^\dagger_\downarrow (\vec r)
f_\downarrow(\vec r) f_\uparrow(\vec r) \Big \} \nonumber \\
&=& \int D \Delta^\dagger D \Delta  D f D f^\dagger
\exp \Big \{ -\int^\beta_0 d\tau \int d\vec r 
\Big \{
\frac{1}{g} \Delta^\dagger(\vec r) \Delta (\vec r) \nonumber \\
&-&\Delta^\dagger(\vec r)  f_\downarrow(\vec r) f_\uparrow(\vec r)
- \Delta(\vec r)  f^\dagger_\uparrow(\vec r) f^\dagger_\downarrow(\vec r)
\Big \}.
\label{eq:HS}
\eea

Now all the fermion terms in Eq. \ref{eq:fermibcs} and 
Eq. \ref{eq:bsfermicnt} become quadratic, we can integrate
out them using the Nambu's representation 
\bea
\int && D f^\dagger D f ~  \exp \Big \{ -\int^\beta_0 d\tau \int d\vec r 
\ \ \ (f^\dagger_\uparrow, f_\downarrow)  \Big \{
\frac{\partial}{\partial \tau}  + (\epsilon(\nabla)  \nonumber \\
&-& \mu) \tau_3
-(\Delta^\dagger(\vec r,\tau ) + \Gamma \phi^\dagger (\vec r, \tau)) \tau_-
-(\Delta(\vec r,\tau ) 
+ \Gamma \nonumber \\
&\times&
 \phi (\vec r, \tau)) \tau_+  \Big \} 
\left(\begin{array}{c}
       f_\uparrow\\
       f^\dagger_\downarrow\\
\end{array}
\right) \nonumber \\
&=&\exp \Big\{ \mbox{tr} \log  \Big \{ \frac{\partial}{\partial \tau}+ (\epsilon(\nabla)-\mu) \tau_3
-(\Re \Delta(\vec r,\tau )  \nonumber \\
&+&  \Gamma ~\Re \phi (\vec r, \tau)) \tau_1 
 -(\Im \Delta(\vec r,\tau ) + \Gamma 
~\Im \phi (\vec r, \tau)) \tau_2\Big \} \Big\}, 
\nonumber \\
\label{eq:integral}
\eea
where $\tau_{1,2,3}$ are the Pauli's matrices in the Nambu's representation,
and $\tau_{+,-}=(\tau_1\pm i\tau_2)/2$.
This leads to the result in Eq.\ref{effective}.

We set the mean field ansatz as
\bea
\Delta(\vec r,\tau)= \bar \Delta +\delta \Delta \ \ \
\phi(\vec r, \tau)= \bar \phi +\delta \phi,
\eea    
where $\bar\Delta$ and $\bar \phi$ are the saddle point value,
while $\delta\Delta$ and $\delta \phi$ are the small
fluctuations.
Then the single particle Green's function reads
\bea
G(\vec p, \tau-\tau^\prime)&=& -
\left( \begin{array} {cc}
{\cal T}  \avg{c_p(\tau) c^\dagger_p(\tau^\prime)}&
{\cal T}  \avg{c_p(\tau) c_{-p}(\tau^\prime)}\\
{\cal T}  \avg{c^\dagger_{-p}(\tau) c^\dagger_p(\tau^\prime)}&
{\cal T}  \avg{c^\dagger_{-p}(\tau) c_{-p}(\tau^\prime)}\\
\end{array} 
\right),\nonumber \\
\eea
where $\cal T$ is the time-order operator.
Its Fourier transforms become
\bea
G(\vec p, ip_n)
&=& \left( \begin{array}{cc}
{\cal G} (p,ip_n)& {\cal F} (p,ip_n) \\
{\cal F}^\dagger (p,ip_n) & -{\cal G} (-p,-ip_n) 
\end{array}
\right) ,
\label{eq:greenfun}
\eea
where the ${\cal G} (p, ip_n)$ and ${\cal F} (p, ip_n)$
are the normal and anomalous Green's functions respectively.
More explicitly, they can be written as
\bea
{\cal G} (p, ip_n) &=&\frac{u_p^2}{i p_n-E_p} + \frac{v^2_p}{ ip_n +E_p}
\nonumber \\
{\cal F} (p, ip_n)
&=& {\cal F^\dagger} (p, ip_n)\nonumber \\
&=& -(u_p v_p) 
\Big\{ \frac{1}{i p_n -E_p} -\frac{1}{i p_n +E_p} \Big\},
\eea
with
\bea
u_p^2=\frac{1}{2}\Big\{ 1+ \frac{\epsilon_p-\mu}{E_p} \Big \}, \ \ \
v_p^2=\frac{1}{2}\Big\{ 1- \frac{\epsilon_p-\mu}{E_p} \Big \}.
\eea
with the dispersion relation 
\bea
E_p^2=(\epsilon_p-\mu)^2 +(\bar \Delta +\Gamma \bar \phi)^2.
\eea

The saddle point equations are determined by the varnishing of 
the first order variations of the effective action Eq. \ref{eq:integral}
over $\delta \Delta$ and $\delta \phi$.
They are
\bea
&&
\frac{\bar \Delta}{g} =\frac{1}{V_L\beta} \sum_{i p_n,p} 
\mbox{tr} \big\{ G(\bar\Delta, \bar \phi; p, i p_n) \tau_- \big\}, \nonumber \\
&&
\frac{\alpha \bar \phi +\lambda |\bar \phi|^2 \bar \phi}{\Gamma}
 =\frac{1}{V_L\beta} \sum_{i p_n,p} 
\mbox{tr} \big\{ G(\bar\Delta, \bar \phi; p, i p_n) \tau_- \big\}.
\eea
After performing the summation over frequency, we arrive at 
Eq. \ref{eq:saddle1}.

\end{document}